\documentclass[aps,prb,twocolumn,superscriptaddress]{revtex4-2}
\usepackage{amsmath}
\usepackage{graphicx}
\usepackage{natbib}
\usepackage{braket}
\usepackage{xcolor}

\begin{document}


\title{Origins of electronic bands in antiferromagnetic topological insulator MnBi\textsubscript{2}Te\textsubscript{4}}


\author{Chenhui Yan}
\author{Sebastian Fernandez-Mulligan}
\affiliation{Pritzker School of Molecular Engineering, University of Chicago, Chicago, Illinois 60637, USA}
\author{Ruobing Mei}
\author{Seng Huat Lee}
\affiliation{Department of Physics, Pennsylvania State University, University Park, State College, PA, 16802, USA}
\author{Nikola Protic}
\author{Rikuto Fukumori}
\affiliation{Pritzker School of Molecular Engineering, University of Chicago, Chicago, Illinois 60637, USA}
\author{Binghai Yan}
\affiliation{Department of Condensed Matter Physics, Weizmann Institute of Science, Rehovot, 7610001, Israel}
\author{Chaoxing Liu}
\author{Zhiqiang Mao}
\affiliation{Department of Physics, Pennsylvania State University, University Park, State College, PA, 16802, USA}
\author{Shuolong Yang}
\email{yangsl@uchicago.edu}
\affiliation{Pritzker School of Molecular Engineering, University of Chicago, Chicago, Illinois 60637, USA}



\date{\today}

\begin{abstract}
Despite the rapid progress in understanding the first intrinsic magnetic topological insulator MnBi$_2$Te$_4$, its electronic structure remains a topic under debates. Here we perform a thorough spectroscopic investigation into the electronic structure of MnBi$_2$Te$_4$ via laser-based angle-resolved photoemission spectroscopy. Through quantitative analysis, we estimate an upper bound of 3 meV for the gap size of the topological surface state. Furthermore, our circular dichroism measurements reveal band chiralities for both the topological surface state and quasi-2D bands, which can be well reproduced in a band hybridization model. A numerical simulation of energy-momentum dispersions based on a four-band model with an additional step potential near the surface provides a promising explanation for the origin of the quasi-2D bands. Our study represents a solid step forward in reconciling the existing controversies in the electronic structure of MnBi$_2$Te$_4$, and provides an important framework to understand the electronic structures of other relevant topological materials MnBi\textsubscript{2n}Te\textsubscript{3n+1}. 
\end{abstract}


\maketitle

\section{Introduction}
Magnetic topological insulators (TIs) host long-range magnetic orders and nontrivial topologies \cite{Tokura2019Magnetic, Kou2015Magnetic}. An energy gap in the topological surface state (TSS) due to the broken time-reversal symmetry \cite{Mong2010Antiferromagnetic, Chen2010Massive} is a prerequisite to realizing exotic many-body phases, such as the Quantum Anomalous Hall (QAH) insulator state \cite{Yu2010Quantized, Chang2013Experimental, Chang2015High-precision} and the Axion insulator state \cite{Mogi2017magnetic, Xiao2018Realization, Liu2020Robust}. As the first intrinsic magnetic TI, MnBi$_2$Te$_4$ exhibits an A-type antiferromagnetic transition near 25 K \cite{Otrokov2019Prediction, Rienks2019Large, Li2019Intrinsic, Li2019Magnetically, Otrokov2019Unique, Zhang2019Topological}, making the material a promising candidate for realizing high temperature topological phases. Encouragingly, the QAH insulator \cite{Deng2020Quantum} and axion insulator \cite{Liu2020Robust} states have been realized in 2D thin layers of MnBi$_2$Te$_4$. However, the electronic band structure of MnBi$_2$Te$_4$ remains a subject under intense debates, even though an enormous amount of work has been done using angle-resolved photoemission spectroscopy (ARPES) \cite{Gong2019Experimental, Li2019Dirac, Chen2019Topological, Hao2019Gapless, Shikin2020Nature, Swatek2020Gapless, Nevola2020Coexistence, Estyunin2020Signatures, Vidal2019Surface}. The first question is whether there is a broken-symmetry gap opening in the TSS. Laser-based ARPES studies resolved a gapless TSS \cite{Li2019Dirac, Chen2019Topological, Hao2019Gapless, Swatek2020Gapless}. Synchrotron-based ARPES studies with high photon energies reported a substantial gap up to ~100 meV \cite{Otrokov2019Prediction, Rienks2019Large}. Transport measurements implied a miniscule gap of 0.64 meV based on the thermal activation behavior of the longitudinal resistance \cite{Deng2020Quantum}. No reconciliation has been attempted on this important fundamental question. Second, the origin of the so-called bulk band splitting across the antiferromagnetic transition remains unclear. A Rashba-type spin texture of the bulk band identified by spin-resolved ARPES \cite{Vidal2019Surface} is inconsistent with the recent two proposals attributing the band splitting to zone folding \cite{Li2019Dirac} or exchange splitting \cite{Chen2019Topological, Hao2019Gapless}. To reconcile these controversies, a comprehensive understanding of the electronic structure of MnBi$_2$Te$_4$ is urgently required.

Here we present an in-depth investigation of the electronic structure of MnBi$_2$Te$_4$ utilizing an ultrahigh resolution laser-ARPES setup with a 6 eV photon energy and a 10 $\mu$m beam waist. We first estimate an upper bound of the gap size in TSS to be 3 meV. More importantly, a circular dichroism (CD) measurement uncovers unique band chiralities of both the TSS and quasi-2D bands. A sign reversal of the chirality on the TSS suggests an anti-crossing behavior. A simple model incorporating band hybridizations reproduces both the experimental band dispersions and CD patterns. Furthermore, we numerically simulate the energy-momentum dispersions based on a four-band model with an additional step potential near the surface. The step potential confines the spin-orbit-coupled quantum well states (QWS) in addition to the TSS, providing a possible microscopic explanation for the quasi-2D bands. Our combined experimental and theoretical studies together elucidate the origins of electronic bands in MnBi$_2$Te$_4$.


\section{Methods}
The MnBi$_2$Te$_4$ samples, grown using the method reported in \cite{lee2020evidence}, were cleaved \emph{in situ} under a base pressure lower than $8 \times 10^{-11}$ mbar. All the ARPES measurements were performed on a Multi-Resolution Spatial and Temporal Engineering Platform (MRSTEP) established at the University of Chicago. This system combines a traditional helium discharge lamp, an 80 MHz Ti:Sa oscillator based 6 eV source, and a 200 kHz ultrafast time-resolved laser setup, which allows us to probe quantum materials holistically in the domains of energy, momentum, space, and time. In this work we mainly invoke the ultrahigh resolution static 6 eV beamline. The overall energy and angular resolutions are better than 4 meV and 0.3°, respectively.  

\section{Results}
\begin{figure}
\includegraphics[width=0.48\textwidth]{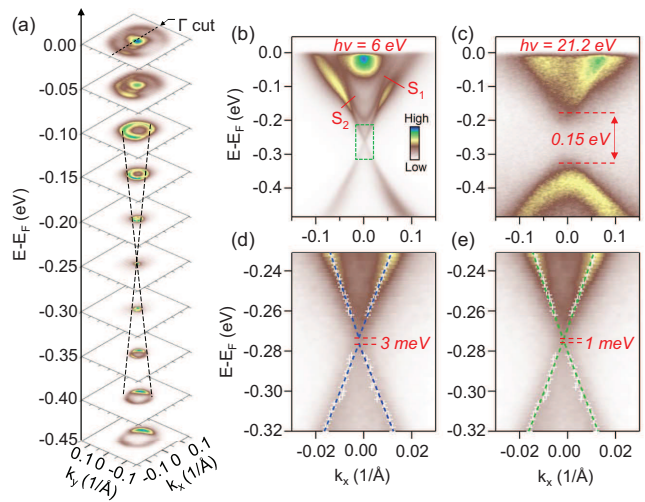}%
\caption{Observation of the topological surface state and quasi-2D bands in MnBi$_2$Te$_4$. (a) ARPES constant energy maps at different binding energies. (b, c) Band dispersions cutting through the $\Gamma$ point taken with a photon energy of (b) 6.04 eV and (c) 21.2 eV, at 15 K. (d, e) Zoomed-in spectra marked by the green dashed box in (b). The overlaid white crossing marks are extracted by fitting momentum distribution curves. The blue and green dashed lines represent fitting curves using (d) linear extrapolation and (e) a more complex model incorporating the band curvatures \cite{Xiong2017Three-dimensional}, respectively.\label{Fig1}}
\end{figure}

The typical constant energy maps and electronic band dispersions of MnBi$_2$Te$_4$ are shown in Fig. \ref{Fig1}. Figure~1(a) shows the constant energy maps at different energies. The outer circular contour at the Fermi level gradually evolves into a single point at -0.25 eV and then to a circle again at lower energies, which is consistent with a cone-like dispersion. This conical TSS dispersion is shown in a cut through the $\Gamma$ point [Fig. 1 (b)]. We extract the TSS dispersion by fitting momentum distribution curves (MDCs) using Lorentzian peaks, and subsequently fit the band dispersion using two models to quantitatively evaluate the gap magnitude. Fitting the dispersion to a linear model leads to a gap size of $3 \pm 1$ meV [Fig. 1(d)]; fitting to a more complex model incorporating potential band curvatures \cite{Xiong2017Three-dimensional} gives rise to a gap size of $1 \pm 1$ meV [Fig. 1(e)]. We emphasize that these values are below our energy resolution (4 meV), and hence only represent our best-effort estimates of the gap. Additionally, these estimates do not exclude the possibility of a zero gap and only serve as upper bounds. Nevertheless, they are consistent with a miniscule gap of 0.64 meV as extracted from a transport study \cite{Deng2020Quantum}. Such a small gap cannot be confidently resolved by ARPES setups with resolutions $\gtrsim 3$ meV, hence reconciling the observation of gapless TSS’s in previous laser-ARPES measurements \cite{Li2019Dirac, Chen2019Topological, Hao2019Gapless, Swatek2020Gapless}. The spectral weight of TSS taken with a high photon energy (21.2 eV) is severely suppressed [Fig.~1(c)], resulting in an apparent gap of ~150 meV, resembling previous ARPES observations with high photon energies \cite{Otrokov2019Prediction, Rienks2019Large, Chen2019Topological}. The measurements with two different photon energies on the same sample overcome the issue of sample quality variations, and suggest that the gap observed using high photon energies $> 20$ eV is likely due to a photon energy dependent matrix element effect instead of antiferromagnetism in MnBi$_2$Te$_4$.

In addition to the TSS, another three bands in MnBi$_2$T$_4$ are clearly observed [Fig. 1(b), Fig. S1]. A broad feature close to the Fermi level with the highest intensity can be attributed to the bulk conduction band. The other two are labeled as “S$_1$”and “S$_2$”, whose linewidths are as narrow as that of the TSS. Furthermore, from previous studies in the literature their dispersions remain the same within experimental uncertainties when the photon energy is substantially increased to $>20$ eV \cite{Vidal2019Surface}. Our observations are in good agreement with previous photon-energy-dependent measurements in the literature \cite{Vidal2019Surface, Ma2021Realization}, implying 2D characters for the “S$_1$” and “S$_2$” bands. 

\begin{figure}
\includegraphics[width=0.48\textwidth]{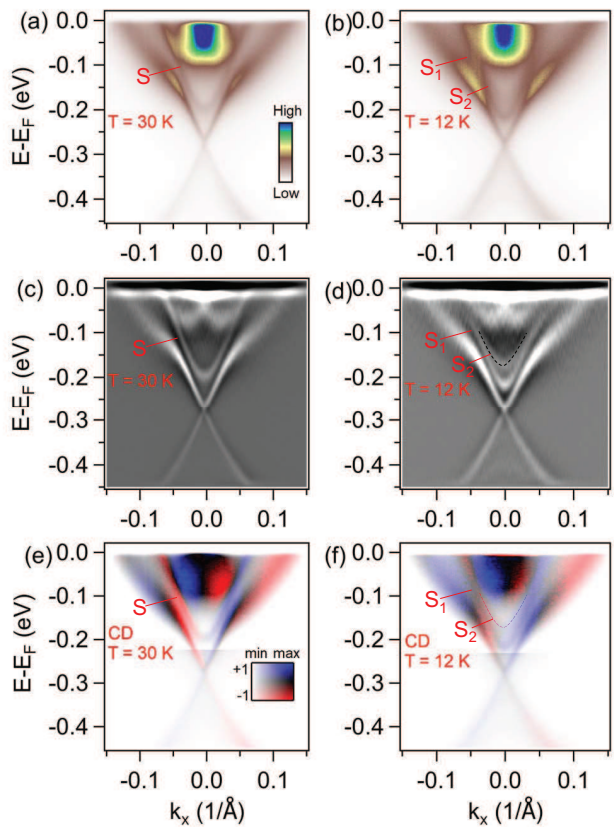}%
\caption{Temperature dependent band dispersions and circular dichroism in MnBi$_2$Te$_4$. (a, b) Band dispersions taken with a photon energy of 6.04 eV at temperatures of (a) 30 K and (b) 12 K. The energy-momentum maps are obtained by adding the spectra taken with LCP and RCP to reduce the polarization matrix element effect. (c, d) Second derivative maps of the spectra in (a) and (b), respectively. The dashed line highlights the “S$_2$” band. (e, f) Circular dichroism maps of the ARPES spectra at (e) 30 K and (f)12 K (the dashed line is a guide to the eyes), respectively. The lower branch of the Dirac cone is plotted using a different color scale as its intensity is much lower than that near the Fermi level. We adopt a two-dimensional color map, where the red-blue color contrast represents circular dichroism and the darkness represents the band intensity \cite{Jozwiak2016Spin-Polarized}.\label{Fig2}}
\end{figure}

Figure \ref{Fig2} displays the splitting of the “S” band at 30 K into the “S$_1$” and “S$_2$” bands at 12 K. This splitting is particularly obvious in the second derivative plots [Fig.~2(c) and 2(d)]. Figure S2 shows a detailed temperature evolution of the band splitting behavior, where the “S$_1$” and “S$_2$” bands merge into the “S” band for temperatures higher than 25 K. The agreement between the splitting onset temperature and the antiferromagnetic transition temperature implies an intimate relationship between these quasi-2D bands and antiferromagnetism. 

Information about chirality is critical in elucidating the origins of the quasi-2D bands. CD in photoemission spectroscopy is derived from an asymmetry in the photoemission spectral weight using left (LCP) and right (RCP) circularly polarized light (Fig. S3). It has been employed to indirectly reveal the spin structure of TSS's in topological insulators, based on spin-dependent photoemission matrix elements \cite{Wang2011Observation, Park2011Orbital-Angular-Momentum, Park2012Chiral, Wang2013Circular, Bahramy2012Emergent}. It is important to note that CD can also detect orbital angular momentum \cite{Park2012Chiral}. In our discussion below, we will not distinguish the chirality introduced by spin or orbital degrees of freedom. We will treat the chirality revealed by CD as simply an intrinsic property of the band structure which helps elucidating the band origins. As shown in Fig. 2(e) and 2(f), the CD pattern on the TSS of MnBi$_2$Te$_4$ below -0.2 eV is consistent with a generic spin-helical structure which has been discovered for all TIs \cite{Wang2011Observation, Park2011Orbital-Angular-Momentum, Park2012Chiral, Wang2013Circular, Bahramy2012Emergent}: it is fully antisymmetric with respect to the Dirac point. Traversing the TSS dispersion towards the Fermi level, we observe a clear “kink” in the dispersion near -0.15 eV [Fig. 2(a) and 2(b)], across which the band sharpness abruptly changes, and the CD notably switches its sign [Fig. 2(e) and 2(f)]. More intriguingly, the “S” band at 30 K exhibits a CD pattern consistent with that on the upper branch of the TSS between the kink energy and the Dirac point energy [Fig. 2(e)]. When the temperature is lowered to 12~K [Fig. 2(f)], the CD pattern on the TSS remains unchanged; the CD pattern on the “S$_1$” band resembles that on the “S” band at higher temperatures; the CD pattern on the “S$_2$” band is not fully distinguishable possibly due to its weak spectral intensity and the proximity to the bulk conduction bands with much higher intensities. Our CD study suggests a nontrivial interplay between the quasi-2D bands and the TSS.

\section{Discussion}
The unique CD pattern in MnBi$_2$Te$_4$ provides strong constraints on the interpretations of the TSS and the quasi-2D bands. The abrupt change in the band sharpness and in the CD pattern of the TSS near the kink energy of -0.15 eV suggests distinct origins for the band segments above and below the kink. Notably, previous ARPES works attributed the quasi-2D bands to trivial bulk states \cite{Li2019Dirac, Chen2019Topological, Estyunin2020Signatures}. Here we emphasize that the photon energy independent dispersions and the band sharpness suggest the 2D characters of the “S”, “S$_1$”, and “S$_2$” bands, which challenge the previous scheme of 3D bulk states \cite{Li2019Dirac, Chen2019Topological, Estyunin2020Signatures}. Moreover, previous studies described two schemes to understand the S$_1$-S$_2$ splitting. In the first scheme, electronic bands are folded along the k$_z$ direction due to unit cell doubling in the antiferromagnetic state, which may give rise to bulk band splitting \cite{Li2019Dirac}. In this band folding scheme, the “S”, “S$_1$”, and “S$_2$” bands are expected to be spin degenerate, which is inconsistent with our observed chiral CD pattern. In the second scheme, electronic bands can be split due to exchange interactions in the antiferromagnetic state \cite{Chen2019Topological, Estyunin2020Signatures}. Yet, in the exchange interaction scheme, the “S” band would be spin degenerate, and the “S$_1$” and “S$_2$” bands would have Zeeman-type spin textures, which are also inconsistent with our observed chiral CD pattern.

\begin{figure}
\includegraphics[width=0.48\textwidth]{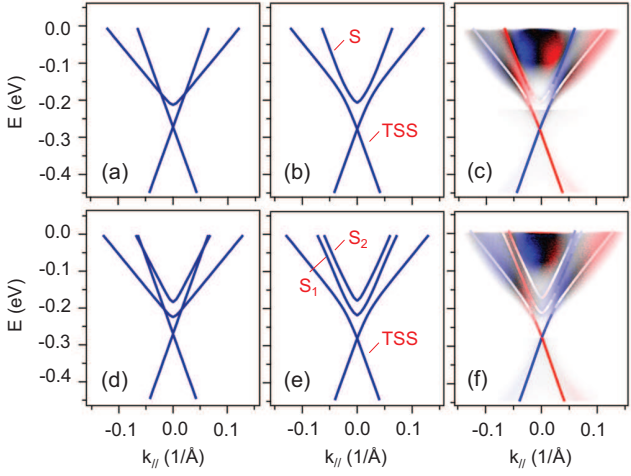}%
\caption{Simulation of hybridization between the TSS and quasi-2D bands. (a) TSS and one quasi-2D band without hybridization. (b) Formed new bands after hybridization. The hybridization energy is 20 meV. (c) The chiral CD pattern of the resulting bands overlaid on the experimental CD map taken at 30 K. (d) TSS and two quasi-2D bands without hybridization. (e) Formed new bands after hybridization. The hybridization energy is 20 meV. (f) The chiral CD pattern of the resulting bands overlaid on the experimental CD map taken at 12 K. The simulated band structures are obtained by solving 2-band or 3-band Hamiltonians with hybridization terms, as discussed in the main text. The simulated CD patterns are obtained by using the proportionate contributions from the unhybridized bands. The proportions are computed in the Hamiltonian diagonalization procedure.\label{Fig3}}
\end{figure}

 Notably, the anti-crossing behavior of the TSS and quasi-2D bands suggests a band hybridization picture. We consider the hybridization of hyperbolic quasi-2D bands with the TSS, for which the exact dispersions are determined by optimizing the agreement with experimental results. This bulk-surface hybridization has been observed in previous ARPES studies on TIs \cite{Kotta2020Spectromicroscopic}. As shown in Fig. 3(a) and 3(b), a hyperbolic quasi-2D band hybridizes with a TSS, which leads to a new hyperbola-like dispersion and a modified TSS with a distinct kink in the TSS band dispersion. Here we use a simple hybridization model, where the hybridization terms appear on the off-diagonal entries of a 2-band Hamiltonian:
\[H=
\begin{bmatrix}
E_{TSS}(k) & \Delta \\
\Delta & E_{2D}(k)
\end{bmatrix}
\]
The hybridization parameter $\Delta$ is chosen to be 20 meV for best agreement with the experimental data (see Fig. S4 and S5 in the Supplemental Material \cite{Yan2021Supplemental}). Importantly, assuming the quasi-2D band possesses a weak CD pattern (20\% polarized), possibly due to chiral orbital angular momentum, the hybridization scheme leads to a CD pattern that resembles the experimental observation in Fig. 2(e). This picture can be easily extended to a 3-band model, where two quasi-2D bands hybridize with the TSS [Fig.~3(e), Fig.~S5], leading to a CD pattern consistent with the low-temperature experiments on MnBi$_2$Te$_4$. 

Despite the simplicity of the hybridization model, the novel features of the TSS and quasi-2D bands can be reasonably reproduced. We thus stress the critical role of hybridization in forming the unique electronic structure of MnBi$_2$Te$_4$. In addition, the mixture of the topologically trivial and nontrivial states due to the hybridization of the TSS and quasi-2D bands could partially account for the miniscule broken-symmetry gap near the Dirac point. In the meantime, this picture does not provide an explanation for the origins of the quasi-2D bands and why there is a band splitting in concomitant with the antiferromagnetic transition. Either band folding or exchange interaction can be incorporated in this theoretical picture. 

\begin{figure}
\includegraphics[width=0.48\textwidth]{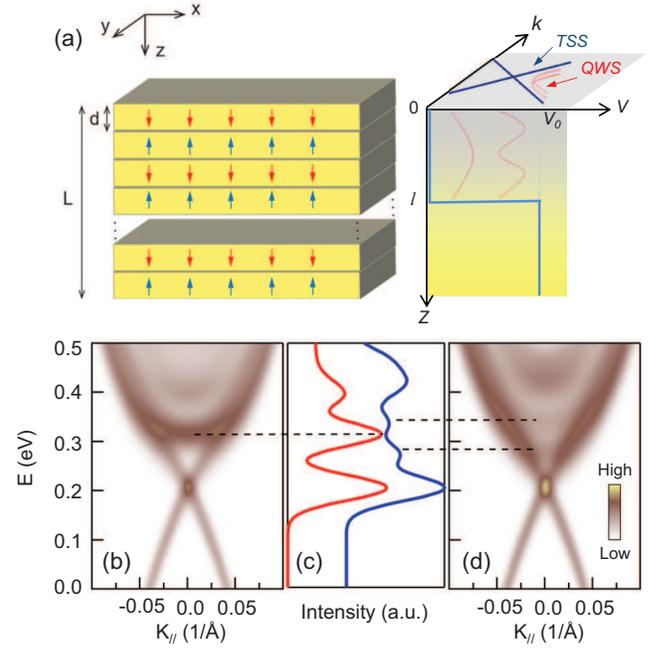}
\caption{Slab calculations of the TSS and quantum-well states (QWS) in MnBi$_2$Te$_4$. (a) Illustration of the model for a MnBi$_2$Te$_4$ slab. The total thickness of the slab is $L$, and the thickness of each monolayer is $d$. $V_0$, $g$, and $\Gamma$ stand for the surface potential, the antiferromagnetic coupling strength, and the scattering rate, respectively. The magnetic moments are represented by red and blue arrows, which flip directions in each neighboring layer. A step potential with length $l$ and depth $V_0$ is invoked to achieve QWS. (b) Density of states (DOS) plot for $V_0 = 0.1$ eV, $g = 0$ eV with momentum-dependent $\Gamma=0.025+2k_{||}^2$. Around 4 surface layers are considered. There is a quasi-2D band between the surface state and conduction bands. (c) Energy distribution curves taken at $k_{||}=0$ (the red and blue ones correspond to the theoretical spectra in (b) and (d), respectively). (d) DOS plot for $V_0 = 0.1$ eV, $g = 0.15$ eV with momentum-dependent $\Gamma=0.025+2k_{||}^2$. The quasi-2D band splits into two sub-bands and the surface state is gapless.\label{Fig4}}
\end{figure}

In an attempt to understand the origins of the quasi-2D bands, we further perform a numerical calculation to simulate a MnBi$_2$Te$_4$ slab based on a four-band model (the details are described in the Supplemental Material \cite{Yan2021Supplemental})\cite{Liu2010Model}. This theoretical construct is a microscopic realization of the aforementioned conceptual framework. Here we attribute the quasi-2D bands to QWS which are confined to the surface due to an additional step potential near the surface, and are strongly influenced by a combination of Rashba splitting and exchange splitting. Notably, 2D QWS with Rashba splitting have been observed in TIs due to a strong band bending near the surface \cite{Bahramy2012Emergent}. With a surface electrostatic potential as shown in Fig.~S6(a), the quasi-2D bands (QWS) with Rashba splitting are realized [Fig. S6(b)]. When antiferromagnetism sets in, the quasi-2D band splits into two subbands due to the exchange splitting [Fig. S6(c)], resembling the experimental temperature dependence of the electronic band structure in MnBi$_2$Te$_4$ (Fig. 2). To reproduce the nearly gapless surface states observed in experiment, we follow proposed schemes in the literature that the time-reversal symmetry is preserved at the top two septuple layers [Fig.S6 (d) and (e)] \cite{Hao2019Gapless}. A momentum-dependent scattering rate is used to mimic the finite band linewidths as observed in ARPES measurements [Fig. 4(b) and 4(d)]. This calculation yields a nearly gapless surface state dispersion hybridized with parabolic quasi-2D bands, which qualitatively explains the experimental observations. We also remark that the comparison of the theoretical spin texture with the experimental CD pattern is more complicated (Fig. S7). The TSS exhibits the same spin texture as that of the lower-energy quasi-2D band (“S$_1$”), yet the spin texture of the higher-energy quasi-2D band (“S$_2$”) is opposite. On the other hand, the experimental CD pattern for the “S$_2$” band in Fig. 2(f) is not clearly resolved, making it difficult to perform the comparison between theory and experiment on this band. Nonetheless, other key features can be qualitatively reproduced by this microscopic calculation.

In summary, by performing a comprehensive spectroscopic investigation into the electronic band structure of MnBi$_2$Te$_4$, we estimate the gap size in the TSS to be at most 1-3 meV and uncover a novel chiral CD pattern of the quasi-2D bands. The CD pattern and band dispersions can be reproduced by a simple model incorporating the hybridization between the TSS and quasi-2D bands. A numerical, microscopic calculation indicates that the quasi-2D bands possibly originate from QWS. These findings highlight the crucial hybridization between the TSS and the quasi-2D bands in MnBi$_2$Te$_4$, providing a conceptual framework for understanding the electronic structures of other relevant topological materials MnBi\textsubscript{2n}Te\textsubscript{3n+1} \cite{Wu2020Distinct, Hu2020Realization, Jo2020Intrinsic, Tian2020Magnetic, Lu2021Half-Magnetic,Vidal2021Orbital,Ma2020hybridization}.

\begin{acknowledgments}
The financial support for sample preparation was provided by the National Science Foundation through the Penn State 2D Crystal Consortium-Materials Innovation Platform (2DCC-MIP) under NSF cooperative agreement DMR-1539916. C. X. L. and R. B. M. acknowledge the support of the U.S. Department of Energy (Grant No. DESC0019064) for the development of the theoretical model.
\end{acknowledgments}

\nocite{Yan2021Supplemental}

\bibliography{originsMBT124}

\end{document}